\documentclass[aps,prl,twocolumn,amsmath,amssymb,nofootinbib,superscriptaddress]{revtex4}
\usepackage[english]{babel}
\usepackage{latexsym}
\usepackage{graphics}
\usepackage{epsfig}
\usepackage{color}
\usepackage{bm}
\usepackage{amsmath}
\usepackage{amssymb}

\usepackage{stackengine}
\stackMath
\newcommand\tenq[2][1]{%
 \def\useanchorwidth{T}%
  \ifnum#1>1%
    \stackon[0pt]{\tenq[\numexpr#1-1\relax]{#2}}{\scriptscriptstyle\sim}%
  \else%
    \stackon[1pt]{#2}{\scriptscriptstyle\sim}%
  \fi%
}

\usepackage[normalem]{ulem}
\usepackage{color}

\begin{document}

\title{Five-Body Efimov Effect and Universal Pentamer in Fermionic Mixtures}
\author{B. Bazak}
\affiliation{IPNO, CNRS/IN2P3, Univ. Paris-Sud, Universit\'e Paris-Saclay, F-91406, Orsay, France }

\author{D.~S.~Petrov}
\affiliation{LPTMS, CNRS, Univ. Paris Sud, Universit\'e Paris-Saclay, F-91405 Orsay, France}
\affiliation{Kavli Institute for Theoretical Physics, University of California, Santa Barbara, Santa Barbara, California 93106, USA}

\date{\today}

\begin{abstract}

We show that four heavy fermions interacting resonantly with a lighter atom (4+1 system) become Efimovian at mass ratio $13.279(2)$, which is smaller than the corresponding 2+1 and 3+1 thresholds. We thus predict the five-body Efimov effect for this system in the regime where any of its subsystem is non-Efimovian. For smaller mass ratios we show the existence and calculate the energy of a universal 4+1 pentamer state, which continues the series of the 2+1 trimer predicted by Kartavtsev and Malykh and 3+1 tetramer discovered by Blume. We also show that the effective-range correction for the light-heavy interaction has a strong effect on all these states and larger effective ranges increase their tendency to bind.

\end{abstract}


\maketitle

Two heavy fermions interacting resonantly with a light atom is an emblematic system exhibiting a transition from the non-Efimovian to Efimovian regime at the mass ratio $M/m=\alpha_c(2,1)=13.607$ \cite{Efimov1973}. The transition clearly demonstrates the interplay of interaction effects and quantum statistics. This can be seen already in the simplest Born-Oppenheimer picture \cite{Efimov1973,Fonseca}; the light atom produces an effective attraction between the heavy atoms proportional to $-1/mR^2$, where $R$ is the separation between the heavy atoms. This attraction competes with the centrifugal barrier $\propto 1/MR^2$ dictated by the fermionic symmetry of the heavy atoms. As one increases $M/m$ above $\alpha_c(2,1)$ the induced attraction wins over the centrifugal barrier and the system becomes Efimovian. It is remarkable that this simple system also exhibits another peculiar effect well inside the non-Efimovian regime. Namely, for a positive heavy-light scattering length $a$ and for $M/m>8.173$ a (non-Efimovian) weakly-bound trimer with unit angular momentum emerges under the atom-dimer scattering threshold \cite{KM}. At smaller mass ratios this trimer turns into a $p$-wave atom-dimer resonance, effects of which have been observed in the $^{40}$K-$^6$Li mixture ($M/m=6.644$) \cite{Rudi}. This manifestly nonperturbative physics has stimulated extensive few- and many-body studies in mass-imbalanced Fermi mixtures \cite{Petrov2003,KM,NishidaSonTan2008,LevinsenPRL2009,PricoupenkoPedri2010,BlumeDaily2010,MathyParishHuse2011,HilfrichHammerJphysB2011,LevinsenPetrovEPJD2011,EndoNaidonUedaFBS2011,CastinTignonePRA2011,LevinsenParish2013,Safavi2013,Mehta2014,EndoCastin2015,KMEPL2016,EndoPRA2016,NaidonEndo2016}.

A natural question is how many identical fermions can be bound by a single light atom? It turns out that three heavy fermions interacting with a light atom become Efimovian for $M/m>\alpha_c(3,1)=13.384$ \cite{CastinPRL2010}, i.e., below the Efimov threshold for the 2+1 subsystem. Blume \cite{BlumePRL2012} has shown that a 3+1 non-Efimovian tetramer with $L^\Pi=1^+$ symmetry emerges below the trimer-atom scattering threshold for $M/m\gtrsim 9.5$ \cite{Remark95}. 
The rapidly increasing configurational space, absence of any small parameter, and need to resolve small energy differences make this problem with more particles significantly more challenging if at all doable with methods used so far \cite{Math}.

In this Letter we solve the 4+1 body problem and show that it is characterized by its proper Efimov threshold at $M/m=\alpha_c(4,1)=13.279(2)$ giving rise to the purely five-body Efimov effect in the range $\alpha_c(4,1)<M/m<\alpha_c(3,1)$. For $M/m<\alpha_c(4,1)$ we find a 4+1 non-Efimovian pentamer that crosses the tetramer-atom threshold at $M/m=9.672(6)$. We argue that considering the heavy-light dimer as a $p$-wave-attractive scattering center for heavy atoms, one builds up the 2+1 trimer, 3+1 tetramer, and 4+1 pentamer by successively filling the $p$ orbitals corresponding to three different projections of the angular momentum. The pentamer has the $L^\Pi=0^-$ symmetry and is the last element of the $p$ shell. This picture is confirmed by our calculation of the energy of the $N$+1-body system in a trap at $a=\infty$. We also include a finite effective range $r_0$ into our analysis and show that the dimer-trimer, trimer-tetramer, and tetramer-pentamer crossings move towards smaller values of $M/m$ with increasing the effective range. This makes the $^{53}$Cr-$^6$Li mixture ($M/m=8.80$) promising for observing the trimer, tetramer, and pentamer phases, transitions among which being realized by tuning the ratio $r_0/a$ and density imbalance.

In order to obtain these results we solve the integral $N$+1-body Skorniakov--Ter-Martirosian (STM) equation by running a stochastic diffusion in the configurational space similar to the diffusion Monte Carlo (DMC) method. Our technique, which we find to work extremely well, combines the advantages of the STM and DMC approaches, it requires no grid and it deals directly with zero-range interactions.

The STM equation was originally derived for the 2+1 mass-balanced problem \cite{STM}. Its generalization to the $N$+1 body problem for negative total energy $E$ in the center-of-mass reference frame in three dimensions reads \cite{Pricoupenko2011}
\begin{widetext}
\begin{equation}\label{STM}
\frac{1}{4\pi}\left(\frac{1}{a}+\frac{r_0\kappa^2}{2}-\kappa \right)F(\mathbf q_1,\ldots, \mathbf q_{N-1})=\int \frac{d^3 q_N}{(2\pi)^3} 
\frac{\sum_{i=1}^{N-1}F(\mathbf q_1,\ldots,\mathbf q_{i-1},\mathbf q_N,\mathbf q_{i+1},\ldots,\mathbf q_{N-1})}
{-\frac{2\mu E}{\hbar^2} +\frac{\mu}{M} \sum_{i=1}^N q_i^2 + \frac{\mu}{m} \left( \sum_{i=1}^{N} \mathbf q_i \right)^2 },
\end{equation}
\end{widetext}
where  $\mu=Mm/(M+m)$ is the reduced mass and $r_0$ is the effective range for the heavy-light interaction. The function $F(\mathbf q_1,\ldots, \mathbf q_{N-1})$ can be considered as the relative wave function of $N-1$ heavy atoms with momenta ${\mathbf q}_1,\ldots,{\mathbf q}_{N-1}$ and a heavy-light pair, the momentum of which equals $-\sum_{i=1}^{N-1} {\mathbf q}_i$ and is thus omitted from the arguments of $F$. More precisely, the coordinate representation of $F$ is $\lim_{{\mathbf r}\rightarrow {\mathbf R}_N}|{\mathbf r}-{\mathbf R}_N|\Psi({\mathbf R}_1,\ldots,{\mathbf R}_N,{\mathbf r})$, where $\Psi$ is the real-space wave function of the $N$+1 system and the singular behavior $\Psi\sim 1/|{\mathbf r}-{\mathbf R}_N|$ is assumed to be true (or extrapolated) down to zero heavy-light distance. In the left-hand side of Eq.~(\ref{STM}) one recognizes the denominator of the heavy-light scattering $t$ matrix at negative collision energy $-\hbar^2\kappa^2/2\mu$ which equals the total energy $E$ minus the kinetic energy of the heavy fermions and the heavy-light pair. Namely, $\kappa=\sqrt{-\frac{2\mu E}{\hbar^2}+\frac{\mu}{M} \sum_{i=1}^{N-1} q_i^2 + \frac{\mu}{M+m} \left( \sum_{i=1}^{N-1} \mathbf q_i \right)^2}$. One can, in principle, take into account higher-order terms in the effective-range expansion, but they compete with the (here neglected) $p$-wave heavy-light and heavy-heavy interaction corrections. It is convenient to regard $E$ as a parameter (together with $r_0$ and $M/m$) and $1/a$ as the eigenvalue to be found. One can then invert the function $a(E)$ in order to find the energy for a given $a$.

Note that $F$ is antisymmetric in all its arguments, and Eq.~(\ref{STM}) does not break this antisymmetry. In addition, Eq.~(\ref{STM}) conserves the angular momentum $L$, its projection, and parity $\Pi$. One can factorize $F$ into a product of a known function $g$ with the same symmetry as $F$ and an unknown function $f$, which depends only on moduli of ${\bf q}_i$ and angles between ${\bf q}_i$ and ${\bf q}_j$. In particular, for the 2+1 problem the relevant (for the Efimov effect and universal trimer) symmetry is $1^{-}$ \cite{KM}, and choosing $m_z=0$ one writes $g({\mathbf q})\propto\hat{\bf z}\cdot {\bf q}$. In the 3+1 case the relevant symmetry is $1^{+}$ and $g({\mathbf q}_1,{\bf q}_2)\propto \hat{\bf z}\cdot {\bf q}_1\times {\mathbf q}_2$ \cite{CastinPRL2010,MoraCRAS}. In the 4+1 case we consider the $0^{-}$ symmetry with $g({\mathbf q}_1,{\mathbf q}_2,{\mathbf q}_3)\propto {\bf q}_1\cdot {\bf q}_2\times {\mathbf q}_3$. Substituting $F=gf$ into Eq.~(\ref{STM}) one obtains STM equations, respectively, for $f(q_1)$ in the 2+1 case, for $f(q_1,q_2,{\bf q}_1\cdot {\bf q}_2)$ in the 3+1 case, and for $f(q_1,q_2,q_3,{\bf q}_2\cdot{\bf q}_3,{\bf q}_3\cdot{\bf q}_1,{\bf q}_1\cdot{\bf q}_2)$ in the 4+1 case. While the first two cases can be solved deterministically by using grid methods, the six-dimensional configurational space of the 4+1 problem is too large for these methods to be sufficiently quantitative.

In order to overcome this problem we develop an exact stochastic method of solving Eq.~(\ref{STM}) inspired by the DMC approach. We note that in the ground state $f$ can be chosen positive. The idea is then to set up a diffusion process in space ${\bf Q}=\{{\mathbf q}_1,\ldots,{\mathbf q}_{N-1}\}$ such that in equilibrium the detailed balance condition for the $3(N-1)$-dimensional element $d{\bf Q}$ is nothing else than Eq.~(\ref{STM}) and the equilibrium density distribution function equals $f({\bf Q})$. 
Let us formally rewrite Eq.~(\ref{STM}) as $f({\bf Q})=\int K({\bf Q},{\bf Q}')f({\bf Q}') d{\bf Q}'$, where the kernel $K({\bf Q},{\bf Q}')>0$ depends on $g({\bf Q})$ \cite{Remarkg} and on parameters $a$, $E$, $r_0$, $M/m$. We search for $a$ at fixed $E$, $r_0$, and $M/m$. The diffusion process is organized as a series of iterations. The input of iteration $i$ is a set of $N_w^{(i)}\gg 1$ walkers with positions ${\bf Q}_1$,\ldots,${\bf Q}_{N_w^{(i)}}$ and a guess for $a$, which we denote $a^{(i)}$. We then calculate walker weights $W_j=\int K({\bf Q},{\bf Q}_j)d{\bf Q}$. Since $W_j$ depend on $a$, we can correct $a$ at this stage such that $\sum_{j=1}^{N_w^{(i)}}W_j$ does not deviate too much from an {\it a priori} set average value $N_w$. We then duplicate each walker on average $W_j$ times and move each new child to a position ${\bf Q}$ drawn from the normalized probability density distribution $K({\bf Q},{\bf Q}_j)/W_j$ \cite{SM}.   
We thus arrive at an updated walker pool with new $N_w^{(i+1)}$ and $a^{(i+1)}$. The process is repeated and, after a thermalization time, which is typically a few tens of iterations, our control parameter $a^{(i)}$ fluctuates around an average value $\langle a \rangle$ and walkers sample $f({\bf Q})$. Strictly speaking this sampling would be exact, if $a^{(i)}$ were the solution. However, fluctuations of $a^{(i)}$ vanish in the limit of large walker number and we have checked that $\langle a \rangle$ converges with increasing $N_w$ \cite{SM}. The efficiency of this algorithm crucially depends on how fast we calculate $W_j$ and sample $K({\bf Q},{\bf Q}_j)/W_j$. For a generic kernel $K({\bf Q},{\bf Q}_j)$ this process would become slow with increasing the dimensionality of ${\bf Q}$, but we benefit from the fact that $K$ is a sum of $N-1$ terms that involve integration (and sampling) only over coordinates of one fermion [see Eq.~(\ref{STM})]. For sufficiently simple $g({\bf Q})$ these tasks are partially analytic and numerically fast \cite{SM}.

\begin{center}
\begin{figure}[ht]
\vskip 0 pt \includegraphics[clip,width=1\columnwidth]{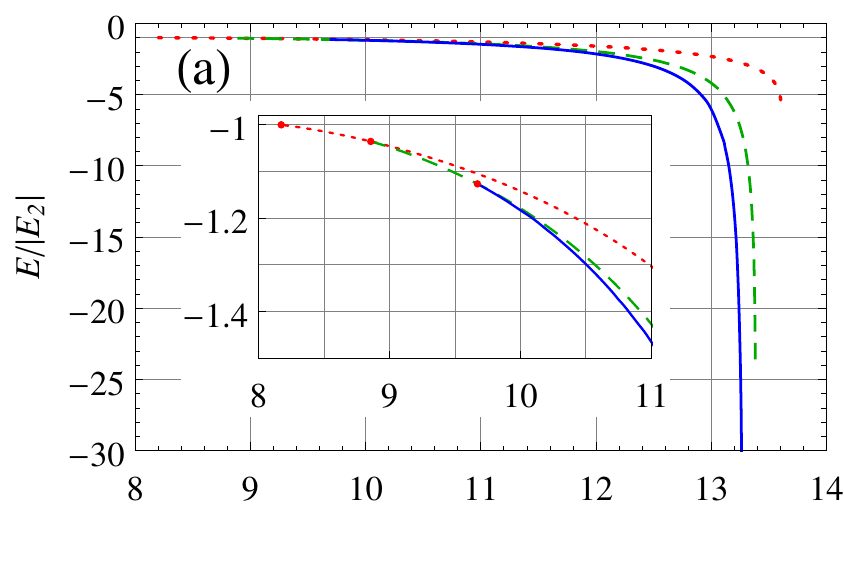}
\vskip -10 pt \includegraphics[clip,width=1\columnwidth]{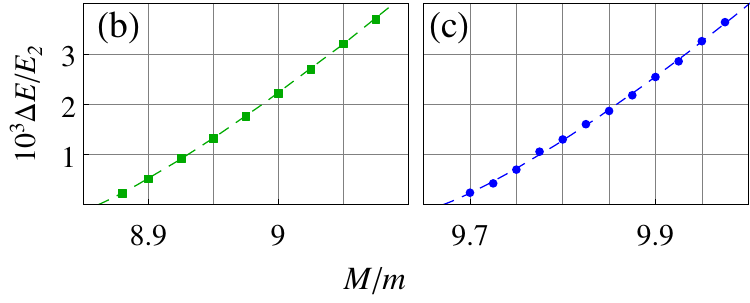}
\caption{
(a) The trimer (dotted red), tetramer (dashed green), and pentamer (solid blue) energies in units of the dimer energy versus $M/m$ (the inset is an enlargement of the crossing region). (b) The trimer-tetramer threshold ($\Delta E=E_4-E_3$). (c) The tetramer-pentamer threshold ($\Delta E=E_5-E_4$). 
}
\label{Fig:Energies}
\end{figure}
\par\end{center}

As the first application of the method let us discuss the energies of bound $N$+1-body states for $a>0$ neglecting the effective range $r_0$. In Fig.~\ref{Fig:Energies}(a) from top to bottom we show the trimer, tetramer, and pentamer energies $E_3$, $E_4$, and $E_5$ in units of the heavy-light dimer energy $E_2=-\hbar^2/2\mu a^2$ as a function of $M/m$ (the inset is an enlargement of the region of their crossings). Our trimer data perfectly reproduce the results of Kartavtsev and Malykh \cite{KM}. As far as the tetramer is concerned, we calculate the trimer-tetramer crossing more precisely than the value $\alpha_{3,4}\approx 9.5$ found by Blume \cite{BlumePRL2012}. Our result $\alpha_{3,4}=8.862(1)$ is obtained by fitting $E_4$ at small but finite $M/m-\alpha_{3,4}>0$ with the threshold law 
$(E_4-E_3)/E_2= 0.011(\alpha_{3,4}-M/m)+0.014(\alpha_{3,4}-M/m)^{3/2}$ 
[see Fig.~\ref{Fig:Energies}(b)]. The branch-cut exponent $3/2$ is related to the density of states in the atom-trimer continuum with unit angular momentum. We find that close to $\alpha_c(3,1)$ the tetramer energy has the threshold behavior 
$E_4/E_2\approx 26(1)-85(2)\sqrt{\alpha_c(3,1)-M/m}$ 
and we confirm the value $\alpha_c(3,1)=13.384$ reported by Castin and co-workers \cite{CastinPRL2010}. 

In the $4+1$ sector we discover two phenomena: the emergence of the universal pentamer state with $0^-$ symmetry and the five-body Efimov effect in the same symmetry channel. We find that the pentamer crosses the tetramer-atom threshold at $\alpha_{4,5}=9.672(6)$ [see Fig.~\ref{Fig:Energies}(c)] and then stays bound up to the five-body Efimov threshold $\alpha_c(4,1)=13.279(2)$ close to which its energy behaves as 
$E_5/E_2\approx 67(3)-310(20)\sqrt{\alpha_c(4,1)-M/m}$
[outside the vertical range in Fig.~\ref{Fig:Energies}(a)]. 

We now discuss the Efimov thresholds for the $N$+1 systems in more detail. The transition from the non-Efimovian to Efimovian regime in these systems is driven by the mass ratio and is associated with a qualitative change in the short-hyperradius behavior of the real-space wave function $\Psi({\mathbf R}_1,\ldots,{\mathbf R}_N,{\mathbf r})$ (see Ref. \cite{CastinPRL2010} and references therein). In brief, one rearranges the relative coordinates into the hyperradius $R\propto \sqrt{m({\bf r}-{\bf C})^2+M\sum_{i=1}^{N}({\bf R}_i-{\bf C})^2}$ (where ${\bf C}$ is the center-of-mass coordinate) and a set of $3N-1$ hyperangles $\hat {\bf R}$. At small $R$ (where $E$ and $1/a$ can be neglected) the hyperradial motion separates from hyperangular degrees of freedom and is then governed by 
\begin{equation}\label{Schr}
\left[-\frac{\partial^2}{\partial R^2}-\frac{3N-1}{R^{2}}\frac{\partial}{\partial R}+\frac{s^2-(3N/2-1)^2}{R^2}\right]\Psi(R)=0,
\end{equation}
where $s^2$ is the hyperangular eigenvalue that depends on $M/m$ (and also on the symmetry of particles and their number, but not on $a$ or $E$). The general solution of Eq.~(\ref{Schr}) is a linear combination of $\Psi_+(R)\propto R^{-3N/2+1+s}$ and $\Psi_-(R)\propto R^{-3N/2+1-s}$.  The case $s^2<0$ ($s=is_0$) corresponds to the Efimovian regime where this linear combination is an oscillating function requiring a three-body parameter to fix the relative phase of $\Psi_+$ and $\Psi_-$. The non-Efimovian regime appears for $s^2>0$ ($s>0$) where, far from few-body resonances, $\Psi(R)$ can be set to be equal to $\Psi_+(R)$ (see, however, Ref. \cite{NishidaSonTan2008,Safavi2013,KMEPL2016}). 

\begin{center}
\begin{figure}[ht]
\vskip 0 pt \includegraphics[clip,width=1\columnwidth]{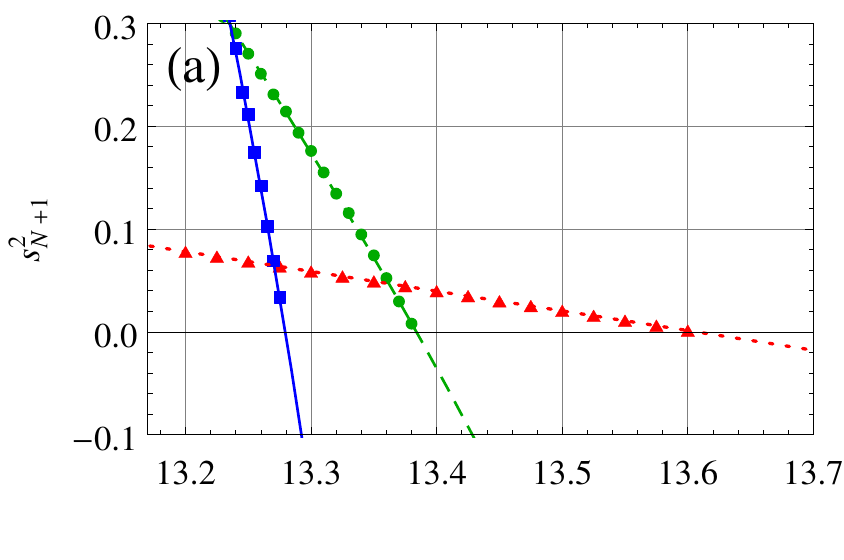}
\vskip -10 pt \includegraphics[clip,width=1\columnwidth]{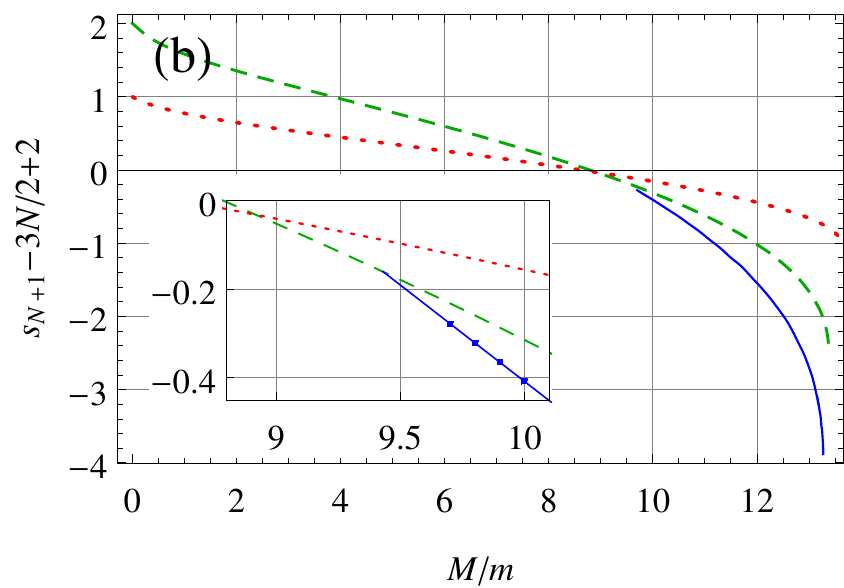}
\caption{
(a) $s_{N+1}^2$ versus $M/m$ close to the $N$+1-body Efimov thresholds for $N=2$ (red triangles), 3 (green circles), and 4 (blue squares) calculated by our stochastic method. The dotted red curve is exact, the dashed green curve is the result of our deterministic grid calculation based on the method of Ref.~\cite{CastinPRL2010}, and the solid blue curve is the linear+quadratic fit to the data. (b) $s_{N+1}-3N/2+2$, related to the energy of the trapped unitary $N$+1 system, versus $M/m$. The color and symbol coding is the same as in (a). The inset shows the crossing region in more detail.}
\label{Fig:Exponents}
\end{figure}
\par\end{center}

In order to determine $s$ by using our method we note that passing from real space to the ${\bf Q}$ space the short-$R$ asymptote $\Psi\propto R^{-3N/2+1+s}$ translates into the large-$Q$ asymptote $F({\bf Q})\propto Q^{-3N/2+1-s}$. While running our algorithm we accumulate statistics for the quantity $\langle |F({\bf Q})|\rangle {}_{\hat{\bf Q}}=\langle f({\bf Q})|g({\bf Q})|\rangle {}_{\hat{\bf Q}}$, 
i.e., every time a walker is found in the bin $(Q,Q+\delta Q)$ we add $|g(\hat{\bf Q})|/\delta Q$ to the bin value if $q_i/Q$ is above a certain small fixed number \cite{SM}. The resulting histogram is fit with the power law $Q^{-3N/2+1-s}$ at large $Q$. In Fig.~\ref{Fig:Exponents}(a) we plot $s^2$ as a function of $M/m$ for the trimer (red triangles), tetramer (green circles), and pentamer (blue squares). In the three-body case this dependence (dotted red) is found exactly by solving a transcendental equation \cite{Petrov2003}. The dashed green curve shows the result of our deterministic grid calculation based on the method of Ref.~\cite{CastinPRL2010}. The solid blue curve is the fit to the pentamer data $s_{4+1}^2= 7.96[\alpha_c(4,1)-M/m] - 25.6 [\alpha_c(4,1)-M/m]^2$ with $\alpha_c(4,1)$ claimed earlier. 

The parameter $s$ in the non-Efimovian case is related to another peculiar universal feature that manifests itself at unitarity ($a=\infty$). In this case, the total energy of the $N$+1 system confined to an isotropic harmonic potential of frequency $\omega$ (same for light and heavy particles) equals $\hbar\omega (s+5/2)$ \cite{Tan,WernerCastin}. In order to compare configurations with different $N$ it is convenient to subtract the energy $3\hbar \omega (N+1)/2$, which is the sum of zero-point single-particle terms. This corresponds to a lattice model with harmonic on-site confinement and vanishing intersite tunneling. In this model the dimer (1+1) and trimer (2+1) energies cross at $M/m=8.6186$, where $s_{2+1}=1$ \cite{Petrov2003}. This means that for $M/m<8.6186$ an on-site heavy-light dimer is formed, but it is energetically favorable for other heavy atoms to be elsewhere. For $M/m>8.6186$ the light atom is able to bind one more heavy atom forming an on-site 2+1 trimer. Increasing the mass ratio further we discover the trimer-tetramer crossing at $M/m=8.918(1)$ and the tetramer-pentamer one at $M/m\approx 9.41(1)<\alpha_{4,5}$ (we have no direct access to the latter crossing point since it is in the region where the uniform-space pentamer is unbound). The curves $s_{N+1}(M/m)-3N/2+2$ are shown in Fig.~\ref{Fig:Exponents}(b). Note that the case $M/m=0$ reduces to the problem of $N$ trapped fermions scattering on a zero-range potential at the trap center. Analyzing the shell structure in this case one obtains $s_{N+1}-3N/2+2 = N-1$ for $N\leq 5$.

\begin{center}
\begin{figure}[ht]
\vskip 0 pt \includegraphics[clip,width=1\columnwidth]{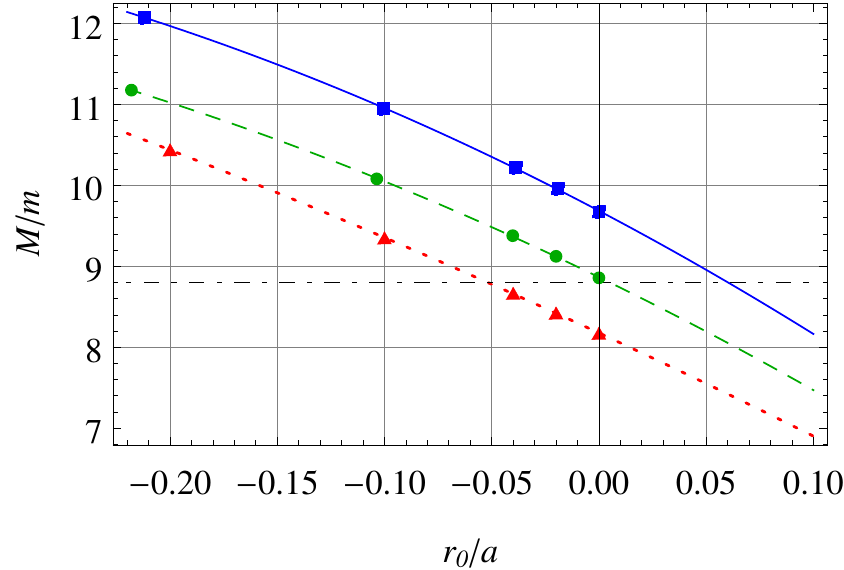}
\caption{
Mass ratios corresponding to the dimer-trimer (dotted red), trimer-tetramer (dashed green), and tetramer-pentamer (solid blue) crossings as a function of $r_0/a$. The solid curves are the linear+quadratic fit to the data. The dash-dotted line is the $^{53}$Cr-$^{6}$Li mass ratio.}
\label{Fig:Crossings}
\end{figure}
\par\end{center}

Let us now discuss effects of finite effective range $r_0$ and assume that the physical ranges of the heavy-heavy and heavy-light potentials are $\sim r_0$. In this case the effective-range expansion involving only $a$ and $r_0$ can be used for calculating few-body observables only up to second order in $r_0$ after which it becomes necessary to include the $p$-wave heavy-heavy and heavy-light contributions (inducing energy shifts $\propto r_0^3$) and the next-order (shape) correction to the $s$-wave heavy-light interaction, which, for sufficiently short-ranged potentials, is of higher order than $r_0^2$ \cite{RemarkPowerCounting,RemarkvdW}. By using our method we calculate $E_{N+1}$ for negative $r_0$ and extrapolate the result to the positive-$r_0$ side limiting ourselves to quadratic terms in $r_0$ \cite{RemarkPositiveRange}. We find that all $N$+1-mers become more bound with increasing $r_0$. The mass ratios corresponding to the dimer-trimer, trimer-tetramer, and tetramer-pentamer crossings as a function of $r_0/a$ are shown in Fig.~\ref{Fig:Crossings}. In view of these findings the $^{53}$Cr-$^6$Li mixture ($M/m=8.80$) emerges as a very promising candidate for observing these bound states; the tetramer turns out to be almost exactly at the threshold for $r_0=0$ and one needs $r_0/a\approx 0.06$ in order to bind the pentamer. Let us point out that, although rather weak, the magnetic dipole-dipole interaction between Cr atoms can become an important factor in determining the energies and crossings of these bound states (cf. Ref. \cite{EndoPRA2016}). These effects require a separate investigation beyond the scope of this Letter.

Our results show that the trimer, tetramer, and pentamer exhibit a remarkable pattern and seem to share a few common features. In particular, they all cross in a rather small window of mass ratios in free space with finite $a>0$ and in a trap at unitarity, their crossings experience an almost parallel shift with $r_0$, etc. In order to understand this phenomenon consider a simplified model of an infinite-mass scattering center (dimer) attracting heavy fermions in the $p$-wave channel. By increasing the attraction one eventually obtains three degenerate bound states that can be filled by heavy fermions. In this model the trimer, tetramer, and pentamer emerge simultaneously and their energies (relative to the dimer one) scale in proportion $1:2:3$. In our case Figs.~\ref{Fig:Energies}(a) and \ref{Fig:Exponents}(b) show a similar behavior demonstrating the shell structure. Based on this model and on the fact that the pentamer closes the shell we conjecture that the 5+1 hexamer and larger clusters of this kind (if bound) should exhibit a qualitatively different behavior and qualitatively different Efimov thresholds (if such thresholds exist). This argument adds importance to the task of calculating the energy and scaling parameter for the 5+1 system since it is definitely too early to directly extrapolate our results to $N\geq 5$ (and eventually to $N\rightarrow \infty$). 
 
We thank G. Astrakharchik, N. Barnea, D. Blume, Y. Castin, C. Greene, U. van Kolck, A. Malykh, D. Phillips, and M. Zaccanti for useful discussions and communications. This research was supported in part by the National Science Foundation under Grant No. NSF PHY-1125915 and received funding from the European Research Council (FR7/2007-2013 Grant Agreement No. 341197), from Chateaubriand Fellowship of the French Embassy in Israel and from the Pazi Fund. We also acknowledge support by the IFRAF Institute.



\clearpage

\renewcommand{\theequation}{S\arabic{equation}}
\renewcommand{\thefigure}{S\arabic{figure}}

\setcounter{equation}{0}
\setcounter{figure}{0}

\onecolumngrid
\center{
\centerline{\underline{\bf SUPPLEMENTAL MATERIAL}}}
\vspace{5mm}
\twocolumngrid

\section{Diffusion method for solving the STM equation}

In this section we describe the diffusion process in detail in the case $N=4$ (pentamer), the other cases being treated in the same manner. We write
\begin{equation}\label{SM:Fgp}
F({\bf q}_1,{\bf q}_2,{\bf q}_3)= g({\mathbf q}_1,{\mathbf q}_2,{\mathbf q}_3)f({\mathbf q}_1,{\mathbf q}_2,{\mathbf q}_3),
\end{equation}
and choose $g$ in the form
\begin{equation}\label{SM:g}
g({\mathbf q}_1,{\mathbf q}_2,{\mathbf q}_3)=\frac{{\bf q}_1\cdot {\bf q}_2\times {\bf q}_3}{|{\bf q}_1\cdot {\bf q}_2\times {\bf q}_3|}\frac{4\pi (q_1^2+q_2^2+q_3^2)^{\frac{\alpha}{2}} q_1^\beta q_2^\beta q_3^\beta}{\kappa_{{\mathbf q}_1,{\mathbf q}_2,{\mathbf q}_3}-\frac{1}{a}-\frac{r_0}{2}\kappa^2_{{\mathbf q}_1,{\mathbf q}_2,{\mathbf q}_3}},
\end{equation}
where $\kappa^2_{{\mathbf q}_1,{\mathbf q}_2,{\mathbf q}_3}=-\frac{2\mu E}{\hbar^2}+\frac{\mu}{M}(q_1^2+q_2^2+q_3^2)+\frac{\mu}{M+m}({\bf q}_1+{\bf q}_2+{\bf q}_3)^2$ and $\alpha$ and $\beta$ are parameters, the choice of which is discussed below.

The function $f({\mathbf q}_1,{\mathbf q}_2,{\mathbf q}_3)$ in Eq.~(\ref{SM:Fgp}) is symmetric with respect to permutations of ${\bf q}_i$ and ${\bf q}_j$. Moreover, it can be written in the form
\begin{equation}\label{SM:p}
f({\mathbf q}_1,{\mathbf q}_2,{\mathbf q}_3)=f(q_1,q_2,q_3,{\bf q}_2\cdot {\bf q}_3,{\bf q}_3\cdot {\bf q}_1,{\bf q}_1\cdot {\bf q}_2),
\end{equation}
which, in particular, means that
\begin{equation}\label{SM:mirror}
f({\mathbf q}_1,{\mathbf q}_2,{\mathbf q}_3)=f(\tenq{\mathbf q}_1,{\mathbf q}_2,{\mathbf q}_3),
\end{equation}
where by $\tenq{\bf q}_i$ we denote the mirror image of ${\bf q}_i$ with respect to the plane spanned by ${\bf q}_j$ and ${\bf q}_k$ ($\{i,j,k\}$ are cyclic permutations of $\{1,2,3\}$). Explicitly, $\tenq{\bf q}_i={\bf q}_i-2({\bf q}_i\cdot \hat{\bf n}_i)\hat{\bf n}_i$, where $\hat{\bf n}_i={\bf q}_j\times {\bf q}_k/|{\bf q}_j\times {\bf q}_k|$. Note that $\kappa$ and $f$ are symmetric and $g$ -- antisymmetric with respect to ${\bf q}_i\rightarrow \tenq{\bf q}_i$. 

Our diffusion process is based on the following. Consider a nine-dimensional element $d^3q_1d^3q_2d^3q_3$ placed at ${\mathbf q}_1,{\mathbf q}_2,{\mathbf q}_3$ and define the distribution function
\begin{widetext}
\begin{equation}\label{SM:Distr}
P_{{\bf q}_1,{\bf q}_2,{\bf q}_3}({\bf q}) =-\frac{{\bf q} \cdot {\bf q}_2\times {\bf q}_3}{|{\bf q} \cdot {\bf q}_2\times {\bf q}_3|}\frac{g({\mathbf q}_1,{\mathbf q}_2,{\mathbf q}_3)}{8\pi^3(q^2+q_2^2+q_3^2)^{\frac{\alpha}{2}}q^\beta q_2^\beta q_3^\beta}\left\{\frac{1}{\kappa^2_{{\mathbf q}_1,{\mathbf q}_2,{\mathbf q}_3}+[{\bf q}+\mu ({\bf q}_1+{\bf q}_2+{\bf q}_3)/m]^2}-({\bf q}_1\rightarrow \tenq{\bf q}_1)\right\},
\end{equation}
which is nowhere negative, and introduce the corresponding normalization integral
\begin{equation}\label{SM:Norm}
W_{{\bf q}_1,{\bf q}_2,{\bf q}_3}=\int d^3 q P_{{\bf q}_1,{\bf q}_2,{\bf q}_3}({\bf q}).
\end{equation}
Assuming that we start with $dN_w$ walkers in $d^3q_1d^3q_2d^3q_3$ we create three groups of new walkers with populations $dN_w W_{{\bf q}_1,{\bf q}_2,{\bf q}_3}$, $dN_w W_{{\bf q}_2,{\bf q}_3,{\bf q}_1}$ and $dN_w W_{{\bf q}_3,{\bf q}_1,{\bf q}_2}$, respectively. Then for each walker in group $i$ we randomly move ${\bf q}_i$ to ${\bf q}$ keeping ${\bf q}_j$ and ${\bf q}_k$ unchanged. Here ${\bf q}$ is drawn from the normalized probability density distribution $P_{{\bf q}_i,{\bf q}_j,{\bf q}_k}({\bf q})/W_{{\bf q}_i,{\bf q}_j,{\bf q}_k}$. 
As a result we obtain $dN_w (W_{{\bf q}_1,{\bf q}_2,{\bf q}_3}+W_{{\bf q}_2,{\bf q}_3,{\bf q}_1}+W_{{\bf q}_3,{\bf q}_1,{\bf q}_2})$ walkers distributed such that only one of their momenta is different from the initial one.

Let us now assume that walkers are initially distributed over the whole space according to the probability density distribution $f({\bf q}_1,{\bf q}_2,{\bf q}_3)$ and consider one such diffusive iteration acting simultaneously over all space elements. Then, the change in the density of walkers equals
\begin{equation}\label{SM:STM}
\delta f({\bf q}_1,{\bf q}_2,{\bf q}_3)=-f({\bf q}_1,{\bf q}_2,{\bf q}_3)+\int d^3 q[P_{{\bf q},{\bf q}_2,{\bf q}_3}({\bf q}_1)f({\bf q},{\bf q}_2,{\bf q}_3)+P_{{\bf q}_1,{\bf q},{\bf q}_3}({\bf q}_2)f({\bf q}_1,{\bf q},{\bf q}_3)+P_{{\bf q}_1,{\bf q}_2,{\bf q}}({\bf q}_3)f({\bf q}_1,{\bf q}_2,{\bf q})].  
\end{equation}
\end{widetext}
The direct substitution of Eqs.~(\ref{SM:Fgp}), (\ref{SM:g}), and (\ref{SM:Distr}) into Eq.~(\ref{SM:STM}) shows that the equilibrium condition $\delta f({\bf q}_1,{\bf q}_2,{\bf q}_3)=0$ leads to the STM Eq.~(1) of the main text.

We model this diffusion process by using a finite number of walkers $N_w^{(i)}$ ($i$ stands for the iteration number), which we keep close to an initially chosen average number $N_w$. For each walker with coordinates ${\bf q}_1,{\bf q}_2,{\bf q}_3$ we create $\lfloor W_{{\bf q}_1,{\bf q}_2,{\bf q}_3}\rfloor$ copies plus another one with probability $W_{{\bf q}_1,{\bf q}_2,{\bf q}_3}-\lfloor W_{{\bf q}_1,{\bf q}_2,{\bf q}_3}\rfloor$, where $\lfloor W \rfloor$ denotes the integer part of $W$. This gives on average $W_{{\bf q}_1,{\bf q}_2,{\bf q}_3}$ copies, the first momentum of which we move to different ${\bf q}$ drawn from $P_{{\bf q}_1,{\bf q}_2,{\bf q}_3}({\bf q})/W_{{\bf q}_1,{\bf q}_2,{\bf q}_3}$. We do the same with the other two momenta.

The total population of walkers, apart from the statistical noise due to the above branching procedure, can be controlled by tuning one of the parameters $E$, $a$, $r_0$, or $M/m$. We typically tune $a$ at fixed $E$, $r_0$, and $M/m$. 
In each iteration we sum $\partial (W_{{\bf q}_1,{\bf q}_2,{\bf q}_3}+W_{{\bf q}_2,{\bf q}_3,{\bf q}_1}+W_{{\bf q}_3,{\bf q}_1,{\bf q}_2})/\partial a$ over the walkers and use it to estimate how strongly we need to change $a$ in order to have $N_w^{(i+1)}$ close to $N_w$. We thus have a sequence of $a^{(i)}$ which fluctuates around an average value. The amplitude of these fluctuations decreases with $N_w$ and $a^{(i)}$ averaged over many iterations converges to the exact $a$ in the limit $N_w \rightarrow \infty$.

\begin{center}
\begin{figure}[ht]
\vskip 0 pt \includegraphics[clip,width=1\columnwidth]{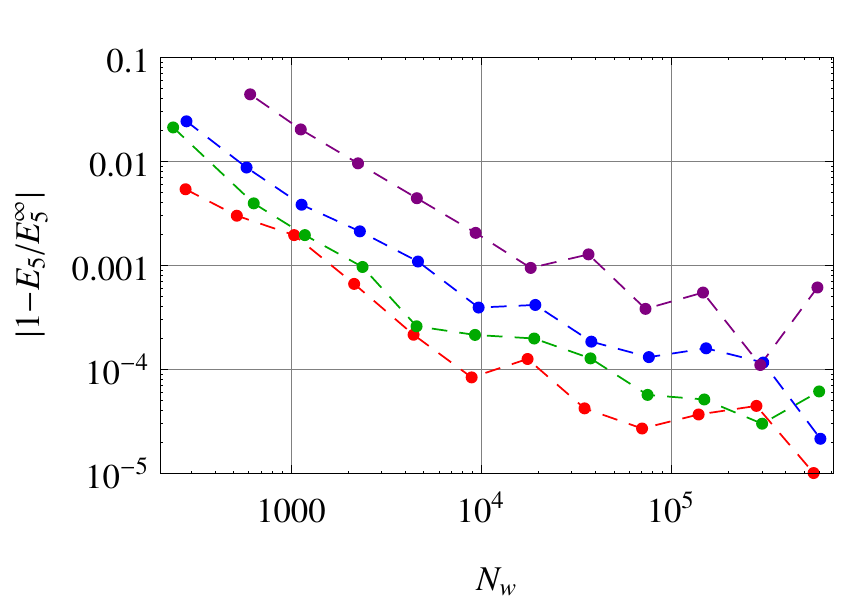}
\caption{
The convergence of the pentamer energy $E_5$ towards $E_5^\infty$ (assumed exact value in the limit $N_w\rightarrow \infty$) with increasing the number of walkers $N_w$ for various $M/m$, from bottom to top: 10 (red), 11 (green), 12 (blue) and 13 (purple).
}
\label{SM:FigConvergence}
\end{figure}
\par\end{center}

In Fig.~\ref{SM:FigConvergence} we show our analysis of the convergence of the pentamer energy with increasing $N_w$ for $M/m = 10$ (red), 11 (green), 12 (blue) and 13 (purple). For these data we use $\alpha = 0$ and $\beta = 2$.

\section{Sampling and normalization}

In this section we outline the sampling procedure for the distribution function (\ref{SM:Distr}) and calculation of the normalization integral (\ref{SM:Norm}). First note that 
\begin{equation}\label{SM:SymmP}
P_{{\bf q}_1,{\bf q}_2,{\bf q}_3}({\bf q}) = P_{{\bf q}_1,{\bf q}_2,{\bf q}_3}(\tenq{\bf q})=P_{\tenq{\bf q}_1,{\bf q}_2,{\bf q}_3}({\bf q}). 
\end{equation}
Therefore, we can restrict ourselves to the domain ${\bf q}_1\cdot {\bf q}_2\times {\bf q}_3>0$ and sample only in the domain ${\bf q}\cdot {\bf q}_2\times {\bf q}_3>0$. We then introduce the reduced distribution function
\begin{equation}\label{SM:DistrReduced}
P^{\rm reduced}_{{\bf q}_1,{\bf q}_2,{\bf q}_3}({\bf q}) =\frac{g({\mathbf q}_1,{\mathbf q}_2,{\mathbf q}_3)}{8\pi^3 q_2^\beta q_3^\beta}\frac{(q^2+q_2^2+q_3^2)^{-\frac{\alpha}{2}}q^{-\beta}}{\kappa^2_{{\mathbf q}_1,{\mathbf q}_2,{\mathbf q}_3}+({\bf q}+{\bf v})^2},
\end{equation}
where ${\bf v}=\mu (\tenq{\bf q}_1+{\bf q}_2+{\bf q}_3)/m$. The distribution function (\ref{SM:DistrReduced}) is, in the chosen domain, larger than $P_{{\bf q}_1,{\bf q}_2,{\bf q}_3}({\bf q})=P^{\rm reduced}_{{\bf q}_1,{\bf q}_2,{\bf q}_3}({\bf q})-P^{\rm reduced}_{{\bf q}_1,{\bf q}_2,{\bf q}_3}(\tenq{\bf q})$. Thus, in order to sample $P_{{\bf q}_1,{\bf q}_2,{\bf q}_3}({\bf q})$ we use the rejection technique. Namely, we draw ${\bf q}$ from (\ref{SM:DistrReduced}) and accept it with the probability $P/P^{\rm reduced}<1$. The sampling of (\ref{SM:DistrReduced}) is realized by using spherical coordinates in which the zenith direction is along ${\bf v}$. Sampling the angles is trivial and the radial coordinate $q$ is distributed according to
\begin{equation}\label{SM:ReducedRadialDistr}
P(q)dq \propto \frac{q^{1-\beta}}{(q^2+q_2^2+q_3^2)^{\frac{\alpha}{2}}}\ln\frac{\kappa^2+v^2+q^2+2vq}{\kappa^2+v^2+q^2-2vq}dq,
\end{equation}
which we sample by using again the rejection method. In particular, in the case $\alpha>2-\beta$ we use the proposal distribution $\propto dq/(\kappa^2+v^2+q^2)$ and the rejection algorithm then relies on the inequality
\begin{equation}\label{SM:Rejection}
\frac{1}{q}\ln\frac{\kappa^2+v^2+q^2+2vq}{\kappa^2+v^2+q^2-2vq}\leq \frac{2\sqrt{\kappa^2+v^2}}{\kappa^2+v^2+q^2}\ln\frac{\sqrt{\kappa^2+v^2}+v}{\sqrt{\kappa^2+v^2}-v}
\end{equation}
and on the fact that $q^{2-\beta}(q^2+q_2^2+q_3^2)^{-\frac{\alpha}{2}}$ is bounded from above. We find that this method works well for all parameters ($\kappa$, $v$, $q_2$, $q_3$, $\alpha$, $\beta$) that we typically deal with.

The normalization integral (\ref{SM:Norm}) is equivalent to integrating $P^{\rm reduced}_{{\bf q}_1,{\bf q}_2,{\bf q}_3}({\bf q})-P^{\rm reduced}_{{\bf q}_1,{\bf q}_2,{\bf q}_3}(\tenq{\bf q})$ over the half space ${\bf q}\cdot {\bf q}_2\times {\bf q}_3>0$. We find it convenient to perform this integration in spherical coordinates with the zenith direction along ${\bf q}_2\times {\bf q}_3$. The angular integrals are analytic and we end up with a one-dimensional integral over $q$ which is numerically fast.
 
\section{Choice of $\alpha$ and $\beta$}

An obvious constraint on possible values of $\alpha$ and $\beta$ is the convergence of the normalization integral (\ref{SM:Norm}). In practice, as we have explained, we require $\alpha>2-\beta$ for sampling convenience. In fact, we find that more strict constraints are dictated by the physics of the problem. As we argue in the main text, for $r_0=0$ the large-${\bf Q}$ asymptote of the function $F$ should be $F({\bf Q})\propto Q^{-3N/2+1-s}$. Then, for $N=4$ the walker distribution function scales at large ${\bf Q}$ as $f({\bf Q})=F({\bf Q})/g({\bf Q})\propto Q^{-4-s-\alpha-3\beta}$ and convergence of $\int f(Q)Q^8dQ$ requires $\alpha+3\beta+s>5$. The same type of convergence condition should also hold for the four- and three-body subsystems of the 4+1 problem. These three conditions can be written simultaneously as
\begin{equation}\label{SM:ConvCond}
\alpha+(N-1)\beta>-s_{N+1}+3N/2-1,
\end{equation}
where $s_{N+1}$ denotes the parameter $s$ for the $N$+1 body (sub)system with $N=2$, 3, and 4 [see Fig.2(b) of the main text]. We should also mention that the condition (\ref{SM:ConvCond}) holds for the pure tetramer and trimer calculations, assuming, respectively, $g({\bf q}_1,{\bf q}_2)\propto (q_1^2+q_2^2)^{\frac{\alpha}{2}}q_1^\beta q_2^\beta/\kappa$ and $g({\bf q}_1)\propto q_1^{\alpha+\beta}/\kappa$.

Trying various combinations of $\alpha$ and $\beta$ we have checked that the result for the energy does not depend on these parameters as long as they satisfy (\ref{SM:ConvCond}) and as long as we use sufficiently large $N_w$. However, for some combinations of $\alpha$ and $\beta$ the calculation of the exponent $s$ should be done with care. This problem arises due to the fact that we are dealing with finite $E$ and $a$. Then the large-$Q$ asymptote of $F({\bf Q})$ at fixed hyperangle $\hat{\bf Q}$ and the large-$Q$ asymptote of the integral $\int F({\bf Q})d\hat{\bf Q}$ do not necessarily coincide. To make this point more clear consider the two-dimensional function $1/[(1+x^2)(1+y^2)]$. For any fixed hyperangle $\theta =\arctan(x/y)$ larger than 0 and smaller than $\pi/2$ this function asymptotes to $\propto 1/\rho^4$ at large hyperradius $\rho=\sqrt{x^2+y^2}$. However, if $\theta=0$ or $\pi/2$, we obtain the $1/\rho^2$ scaling. Moreover, if we integrate $1/[(1+x^2)(1+y^2)]$ over $\theta$, we obtain yet another power law $\propto 1/\rho^3$. 

Clearly, the power-law scaling that one obtains from a function of the type $(x^2+y^2)^{\frac{\alpha}{2}}x^\beta y^\beta/[(1+x^2)(1+y^2)]$ depends on $\alpha$, $\beta$, and on the exact limiting procedure. A possible solution of this problem is to restrict $\alpha$ and $\beta$ such that the total integral over the hyperangles is properly behaved. In the particular example just considered we have to choose $\beta$ such that the integral $x^\beta/(1+x^2)$ diverges at large $x$, i.e., $\beta>1$, thus effectively depreciating the role of the small-$x$ region. A much simpler solution is to restrict the hyperangular integration to the region $\epsilon<\theta<\pi/2-\epsilon$. This is what we do in the actual calculations. Namely, when we gather statistics on walkers in the interval $(Q,Q+\delta Q)$, we update the bin value only if $q_i/Q>\epsilon$.  


\begin{thebibliography}{99}
\bibitem{RemarkPositiveRange}
Our method cannot be directly applied in the case $r_0>0$ since the prefactor $1/a+r_0\kappa^2/2-\kappa$ in Eq.~(\ref{STM}) has a node at large $\kappa\sim 1/r_0$. The problem can be avoided by introducing a momentum cutoff, beyond-effective-range terms, or, with the same accuracy, by extrapolating from the negative-$r_0$ side.
\end{thebibliography}
\end{document}